\documentclass[traditabstract, longauth]{aa} 
\usepackage{graphicx}
\usepackage{txfonts}
\usepackage{natbib,twoopt}
\usepackage{longtable}

\def\ms{\hbox{\,m\,s$^{-1}$}}         
\def\m2s2{\hbox{\,m$^{2}$\,s$^{-2}$}} 
\def\kms{\hbox{\,km\,s$^{-1}$}}       
\def\Msun{\hbox{$\mathrm{M}_{\odot}$}}             
\def\Rsun{\hbox{$\mathrm{R}_{\odot}$}}
\def\Mjup{\hbox{$\mathrm{M}_{\rm Jup}$}}
\def\Rjup{\hbox{$\mathrm{R}_{\rm Jup}$}}
\def\Mearth{\hbox{$\mathrm{M}_{\oplus}$}}             
\def\Rearth{\hbox{$\mathrm{R}_{\oplus}$}}

\def\mp{M_{\rm p}}
\def\rp{R_{\rm p}}

\begin{document}

\title{A deeper view of the \object{CoRoT-9} planetary system} 
\subtitle{A small non-zero eccentricity for \object{CoRoT-9b} \\
likely generated by planet-planet scattering}
\titlerunning{A deeper view of the CoRoT-9 planetary system}
\authorrunning{Bonomo et al. 2017}

\author{
A. S. Bonomo \inst{\ref{oato}} 
\and G. H\'ebrard \inst{\ref{iap}, \ref{ohp}}
\and S. N. Raymond\inst{\ref{lab}}
\and F. Bouchy\inst{\ref{geneva}}
\and A. Lecavelier des Etangs\inst{\ref{iap}}
\and P. Bord\'e\inst{\ref{lab}} 
\and S. Aigrain \inst{\ref{oxford}}
\and J.-M. Almenara\inst{\ref{geneva}} 
\and R. Alonso\inst{\ref{iac}, \ref{UnivCan}} 
\and J. Cabrera\inst{\ref{dlr}}
\and Sz. Csizmadia\inst{\ref{dlr}}
\and C. Damiani\inst{\ref{ias}}
\and H. J. Deeg\inst{\ref{iac},\ref{UnivCan}}
\and M. Deleuil\inst{\ref{lam}}
\and R. F. D\'iaz\inst{\ref{unibuenos}, \ref{conicet}}
\and A. Erikson\inst{\ref{dlr}}
\and M. Fridlund\inst{\ref{leiden}, \ref{onsala}}
\and D. Gandolfi\inst{\ref{unito}}
\and E. Guenther\inst{\ref{tls}}
\and T. Guillot\inst{\ref{oca}}
\and A. Hatzes\inst{\ref{tls}}
\and A. Izidoro\inst{\ref{unesp}, \ref{lab}}
\and C. Lovis\inst{\ref{geneva}}
\and C. Moutou\inst{\ref{cfht}, \ref{lam}} 
\and M. Ollivier\inst{\ref{ias}, \ref{lesia}}
\and M. P\"atzold\inst{\ref{koln}}
\and H. Rauer\inst{\ref{dlr}, \ref{ZAA}}
\and D. Rouan\inst{\ref{lesia}}
\and A. Santerne\inst{\ref{lam}}
\and J. Schneider\inst{\ref{luth}}
}


\institute{
INAF - Osservatorio Astrofisico di Torino, via Osservatorio 20, 10025 Pino Torinese, Italy\label{oato}
\and Institut d'Astrophysique de Paris, CNRS, UMR~7095 \& Sorbonne Universit\'es, UPMC Paris 6, 98 bis bd Arago, 75014 Paris, France\label{iap}
\and Observatoire de Haute Provence, Universit\'e Aix-Marseille \& CNRS, F-04870 St.~Michel l'Observatoire, France\label{ohp}
\and Laboratoire d'Astrophysique de Bordeaux, Universit\'e de Bordeaux, CNRS, B18N, all\'ee Geoffroy Saint-Hilaire, F-33615 Pessac, France\label{lab}
\and Observatoire de l'Universit\'e de Gen\`eve, 51 chemin des Maillettes, 1290 Sauverny, Switzerland\label{geneva} 
\and Department of Physics, Denys Wilkinson Building Keble Road, Oxford, OX1 3RH\label{oxford}
\and Instituto de Astrofisica de Canarias, E-38205 La Laguna, Tenerife, Spain\label{iac}
\and Universidad de La Laguna, Dept. de Astrof\'\i sica, E-38200 La Laguna, Tenerife, Spain\label{UnivCan}
\and Institute of Planetary Research, German Aerospace Center, Rutherfordstrasse 2, 12489 Berlin, Germany\label{dlr}
\and Institut d'Astrophysique Spatiale, UMR 8617, CNRS - Universit\'e de Paris-Sud - Universit\'e Paris-Saclay - B\^at. 121, 91405 Orsay, France\label{ias}
\and Aix Marseille Universit\'e, CNRS, LAM (Laboratoire d'Astrophysique de Marseille) UMR 7326, 13388, Marseille, France\label{lam}
\and Universidad de Buenos Aires, Facultad de Ciencias Exactas y Naturales. Buenos Aires, Argentina\label{unibuenos}
\and CONICET - Universidad de Buenos Aires. Instituto de Astronom\'ia y F\'isica del Espacio (IAFE). Buenos Aires, Argentina\label{conicet}
\and Leiden Observatory, University of Leiden, P.O. Box 9513, NL-2300 RA, Leiden, The Netherlands\label{leiden}
\and Department of Earth and Space Sciences, Chalmers University of Technology, Onsala Space Observatory, SE - 439 92 Onsala\label{onsala}
\and Dipartimento di Fisica, Universit\`a di Torino, via P. Giuria 1, I-10125 Torino, Italy\label{unito}
\and Th\"uringer Landessternwarte, Sternwarte 5, Tautenburg 5, D-07778 Tautenburg, Germany\label{tls}
\and Observatoire de la C\^ote d'Azur, Laboratoire Cassiop\'ee, BP 4229, 06304 Nice Cedex 4, France\label{oca}
\and UNESP, Univ. Estadual Paulista - Grupo de Din\^amica Orbital \& Planetologia, Guaratinguet\'a, CEP 12.516-410, S\~ao Paulo, Brazil\label{unesp}
\and CNRS, CFHT Corporation, 65-1238 Mamalahoa Hwy, Kamuela, HI 96743, USA\label{cfht}
\and LESIA, Obs de Paris, Place J. Janssen, 92195 Meudon cedex, France\label{lesia}
\and Rheinisches Institut f\"ur Umweltforschung an der Universit\"at zu K\"oln, Aachener Strasse 209, 50931, Germany\label{koln} 
\and Center for Astronomy and Astrophysics, TU Berlin, Hardenbergstr. 36, 10623 Berlin, Germany\label{ZAA}
\and LUTH, Observatoire de Paris, CNRS, Universit\'e Paris Diderot; 5 place Jules Janssen, 92195 Meudon, France \label{luth}
}

\date{Received 15 February 2017 / Accepted 17 March 2017}

\offprints{\email{bonomo@oato.inaf.it}}

\abstract{
\object{CoRoT-9b} is one of the rare long-period ($P=95.3$~days) transiting giant planets with a measured mass known to date.  
We present a new analysis of the \object{CoRoT-9} system based on five years of radial-velocity (RV) monitoring with HARPS 
and three new space-based transits observed with CoRoT and Spitzer. 
Combining our new data with already published measurements we redetermine the CoRoT-9 system parameters and find good agreement with the published values. We uncover a higher significance for the small but non-zero eccentricity of CoRoT-9b ($e=0.133^{+0.042}_{-0.037}$) and find no evidence for additional planets in the system.  We use simulations of planet-planet scattering to show that the eccentricity of CoRoT-9b may have been generated by an instability in which a $\sim 50~\Mearth$ planet was ejected from the system. 
This scattering would not have produced a spin-orbit misalignment, so we predict that the CoRoT-9b orbit should lie within a few degrees of the initial plane of the protoplanetary disk. As a consequence, any significant stellar obliquity would indicate that the disk was primordially tilted. }

\keywords{Planetary systems -- Techniques: photometric -- Techniques: radial velocities -- Stars: individual: CoRoT-9}

\maketitle

%

\section{Introduction} 
As of February 2017, only 27 warm Jupiters (WJs), 
defined as giant planets ($\mp > 0.1~\Mjup$) with 
orbital distance $0.1 < a < 1$~AU \citep[e.g.][]{dawson14},
are known to transit in front of their host stars and to
have a measured mass better than $3\sigma$\footnote{data from http://exoplanetarchive.ipac.caltech.edu and http://exoplanet.eu.}. 
All except four of these WJs were discovered  by space-based missions as, in general, 
only these surveys provide photometric time series of sufficient precision, length, and sampling consistency to detect them.
Indeed, the first WJ detected via the transit method 
was discovered by the CoRoT space telescope;  this object, called \object{CoRoT-9b},  orbits a non-active G3V star 
with orbital period $P=95.3$~d and semimajor axis $a=0.41$~AU, and 
has a mass of $0.84\pm0.07~\Mjup$ and a radius of $1.05\pm0.04~\Rjup$ 
\citep[][hereafter D10]{2010Natur.464..384D}. 
Thanks to the \emph{Kepler} space mission, many more WJ candidates 
could be found and, for some of them, radial-velocity (RV) follow up 
\citep[e.g.][and references therein]{2012A&A...545A..76S, 2016A&A...587A..64S} 
and/or analysis of transit time variations 
\citep[e.g.][]{2012ApJ...761..163D, 2014A&A...571A..38B, 2015A&A...573A.124B} have
allowed both to unveil their planetary nature and determine their mass and hence their densities.
Specifically, 18 of the aforementioned 27 transiting WJs were discovered by \emph{Kepler}. 

The discovery and characterisation of transiting WJs are of great importance 
to better understand the  internal structure, formation, and 
evolution of giant planets. For instance, the mass-radius relation of giant planets at orbital distances $a > 0.1$~AU
should not be affected by stellar heating, which is likely related to the inflation mechanism responsible 
for the large radii of several hot Jupiters 
\citep{2011A&A...532A..79S, 2011ApJS..197...12D, 2015A&A...575L..15S}.
Therefore, WJs are not expected to be inflated unless other processes are at work. 

The formation and orbital evolution of WJs is currently a very interesting issue of debate. 
None of the processes that have been invoked to explain the population of close-in giant exoplanets obviously applies to WJs.  
Inward type~II migration \citep[e.g.][]{1996Natur.380..606L} halted by disk photoevaporation 
may produce WJs \citep[e.g.][]{alexander12, 2012A&A...541A..97M}, but this migration cannot explain the high eccentricities of 
many of them \citep{2017arXiv170400373B}
given that migration in the disk only tends to damp non-zero eccentricities \citep[e.g.][]{kley12}. 
In addition, planet-planet scattering becomes less effective closer to the star \citep{2014ApJ...786..101P}.  
Based on a population-synthesis study, \citet{petrovich16} found that $\sim20\%$ of WJs
may have migrated through high-eccentricity migration \citep[e.g.][]{1996Sci...274..954R}, in particular  
through the high-eccentricity phase of secular oscillations excited by an outer companion in an 
eccentric and/or mutually inclined orbit \citep[see also][]{2011ApJ...735..109W}. 
On the contrary, \citet{2017MNRAS.464..688H} were not able to produce any WJ
from secular evolution in multi-planet systems with three to five giant planets and suggested
that WJs either underwent disk migration or formed in situ. In situ formation is proposed by 
\citet{2016ApJ...825...98H} as the most likely mechanism to form multi-planet 
systems with WJs flanked by close and smaller companions, such as those observed by \emph{Kepler}. 
The rate of occurrence of WJs in such systems may be relatively high, up to $\sim50\%$,
although it is highly uncertain at the moment \citep{2016ApJ...825...98H}. 
The same authors argued that the WJs without known close companions 
might represent a distinct population that formed and migrated differently from 
the former population (e.g. WJs in compact multi-planet systems). 

Yet, it is also possible that WJs with no close-in small companions are simply the innermost planetary cores of the system 
that grew into gas giants and migrated inward. 
The birth and migration of a gas giant play a crucial role in the evolution and dynamics of a young planetary system. 
In this context, two aspects can help to explain the lack of (detected) inner companions in these specific systems. 
The first aspect is that a (forming) gas giant stops the radial flux of small planetesimal and pebbles 
drifting inward due to gas drag \citep{2014A&A...572A..35L}.
This could cause the region inside the orbit of the putative growing and migrating WJ to be too  low mass 
to support, for instance, the formation of any planet larger than the Earth. 
Secondly, a gas giant formed from the innermost planetary core in the system also acts as an efficient dynamical barrier 
to additional inward-migrating planetary cores (or planets) formed on outer orbits \citep{2015ApJ...800L..22I}. 
Thus, inward-migrating planets from external parts of the disk typically cannot make their full way inward to the innermost regions. 
They tend to be captured in external mean motion resonances with the gas giant. 
This could also help to explain the lack of inner companions in these systems.  
On the other hand, one should naturally expect that outer companions to WJs should be common in this context. 
However, the subsequent dynamical evolution of these planetary systems 
post-gas dispersal (e.g. occurrence of dynamical instabilities or not) is determinant in setting the real destiny of these planets.

Monitoring of planetary systems containing WJs 
that are not flanked by close and small companions,
with RV and/or (Gaia) astrometric measurements 
as well as observations of adaptive optics imaging are crucial to
i) search for outer (planetary or stellar) companions; 
ii) provide information on whether these exterior companions may have triggered 
high-eccentricity migration \citep[e.g.][]{2016ApJ...821...89B}; and 
iii) detect significant orbital eccentricities that could be an imprint of secular chaos or planet-planet scattering possibly 
occurring after early disk\ migration \citep[e.g.][]{2010A&A...514L...4M}.

CoRoT-9b belongs to the class of WJs with no detected close transiting companions. 
In this work we present photometric follow-up with CoRoT and Spitzer (Sect.~\ref{photometric_data}) 
and spectroscopic monitoring with HARPS for a total time span of almost five years (Sect.~\ref{RV_data}). 
With a combined Bayesian analysis of photometric and RV data (Sect.~\ref{data_analysis}), 
we redetermine the system parameters and uncover a higher significance 
for the small eccentricity of CoRoT-9b (Sect.~\ref{system_parameters}). 
We find no evidence for additional planetary companions in the system with the gathered 
data (Sect.~\ref{CoRoT-9b_alone}). Finally, we investigate several scenarios 
for the possible formation and migration of CoRoT-9b (Sect.~\ref{evolutionary_scenarios})
and carry out planet-planet scattering simulations to reconstruct the dynamical history
of the \object{CoRoT-9} planetary system that best matches the observational constraints  
(Sect.~\ref{scattering_simulations}).

\section{Data}
\subsection{Photometric data}
\label{photometric_data}
In addition to one full and one partial transit of CoRoT-9b observed by the CoRoT satellite in 2008 (D10), 
here we make use of two more transits: the first was observed by Spitzer on 18 June 2010 and 
the second on 4 July 2011 simultaneously by Spitzer and CoRoT (see Fig.~\ref{fig_transits}). 
The temporal sampling of CoRoT and Spitzer data is 32 and 31~s, respectively. 

CoRoT photometry with the imagette pipeline (e.g. \citealt{2014A&A...569A..74B}) was obtained 
during the CoRoT SRc03 pointing\footnote{data available at the CoRoT archive: http://idoc-corot.ias.u-psud.fr}, 
which lasted only five days and was dedicated to the observation of the CoRoT-9b transit.
Flux contamination by background stars in the CoRoT mask \citep{2006ESASP1306..293L, 2009A&A...506..343D} 
was estimated to be very low, that is $2.5\%$, by D10 for the data acquired in 2008; we adopted the same value.
However, the imagette pipeline does not permit us to estimate such a contamination for the transit observations in 2011 as well;
hence we included a dilution factor as an additional free parameter in the transit fitting (Sect.~\ref{data_analysis}).

Both Spitzer observations were secured with the Channel~2 at 4.5~$\mu$m of the IRAC 
camera \citep{2004ApJS..154...10F}. These infrared observations and their reduction are described in 
\citet{lecavelier_etal_2017} 
who present a search for signatures of rings and satellites 
around CoRoT-9b. Here we use those Spitzer light curves to refine the parameters of CoRoT-9b and its host star. 
Briefly, we used the basic calibrated data files of each image as produced by the IRAC pipeline, 
then corrected them for the so-called pixel-phase effect, which is the oscillation of the measured flux due to the 
Spitzer jitter and the detector intra-pixel sensitivity variations.
Despite this correction, the 2010 Spitzer transit light curve shows two residual systematic effects 
\citep{lecavelier_etal_2017}: first, it presents a ``bump'' near the middle of the transit, which 
is unlikely to be due to the crossing of a large starspot by the planet given that the host star is very quiet  
(no activity features are seen in the CoRoT light curve); 
secondly, the transit is significantly deeper than that 
observed in 2011 by $\sim 13\%$ (see Fig.~\ref{fig_transits} and Sect.~\ref{system_parameters}). 
This larger depth cannot be attributed to unocculted starspots because it would imply 
a spot filling factor of $\sim 40\%$\footnote{estimated 
from Eq.~7 in \citet{2012A&A...539A.140B} by considering an early G-type star and the 
4.5~$\mu \rm m$ Spitzer band (see their Tables~1 and 2).} that is again unrealistic for the CoRoT-9 low activity level. 
To overcome these effects, we excluded from our analysis the data points in the bump and, 
similarly to the 2011 CoRoT transit, we considered a dilution factor for the 2010 Spitzer data 
to account for its larger depth (see Sect.~\ref{data_analysis}). By doing so, we substantially rely on the 2011 Spitzer 
transit for the determination of the planetary radius at $4.5~\mu \rm m$ because 
this transit does not show any feature that might be related to residual systematic effects.

CoRoT and Spitzer transit light curves were normalised following \citet{2015A&A...575A..85B};  
we excluded the partial CoRoT transit for the determination of system parameters (Sect.~\ref{data_analysis}) 
because of a possible non-optimal normalisation due to the lack of egress data, but we used it for the computation of 
transit timing variations (TTVs) (Sect.~\ref{data_analysis} and \ref{CoRoT-9b_alone}). 
Correlated noise on an hourly timescale in each light curve was estimated 
as in \citet{2009A&A...506..425A} and \citet{2012A&A...547A.110B} 
but was found to be practically negligible for all the transits.
After subtracting the transit model (Sect.~\ref{system_parameters}), 
the CoRoT data have an r.m.s. of $\sim2.8 \cdot 10^{-3}$ in units of relative flux
while the Spitzer measurements show a higher r.m.s. of $\sim4.5 \cdot 10^{-3}$;
there is no significant difference in the r.m.s. among the CoRoT transits or between the two Spitzer transits.

We also searched for additional transit signals in the CoRoT data with the LAM pipeline (\citealt{2012A&A...547A.110B};  
see also \citealt{2009A&A...495..647B}) after removing the CoRoT-9b transits, but found none 
(see Sect.~\ref{CoRoT-9b_alone}).

\subsection{Radial-velocity data}
\label{RV_data}
We obtained 28 radial-velocity observations of CoRoT-9 between September 2008 
and August 2013 with the HARPS fibre-fed spectrograph \citep{2003Msngr.114...20M}
at the 3.6 m ESO telescope in La Silla, Chile (programme 184.C-0639). 
The resolution power is 115,000. Depending on the observations, exposure times 
range between 40 and 60~min and provide signal-to-noise ratios between 11 and 24 per pixel at 550~nm.
All the observations were gathered with the high-accuracy mode (HAM) of HARPS except that at $\rm BJD_{UTC}$=2454766.51,
which was secured in high-efficiency (EGGS) mode; we decided to keep this observation in 
our analysis as it shows no significant drift in radial velocity.

The spectra were extracted using the HARPS pipeline and the radial velocities 
were measured from the weighted cross-correlation with a numerical mask
\citep{1996A&AS..119..373B, 2002A&A...388..632P}. We tested masks 
representative of F0, G2, and K5 stars. 
The bluest of the 68 HARPS spectral orders are noisy for that relatively faint ($V=13.7$) star so 
we adjusted the number of orders used in the cross-correlation. The solution
we adopt is the cross-correlation performed with the K5-type mask with the exclusion of the 
10 first blue orders. We chose that configuration because it minimises the dispersion
of the RV residuals after the Keplerian fit. Other configurations do not provide 
significantly different system parameters, but their residual dispersions are larger. 
Following the method presented in \citet{2010A&A...520A..65B}, 
moonlight contamination was corrected for 10 observations using the second optical-fibre 
aperture targeted on the sky. The corrections are of the order of $10~\ms$ or less, except 
for the two most polluted exposures where they are between 25 and $50~\ms$.

The RV measurements are reported in Table~\ref{table_rv} and shown
in Fig.~\ref{fig_RV}. 
These measurements show variations in phase with the transit ephemeris derived from CoRoT and Spitzer 
photometry. The bisector spans of the cross-correlation function show neither variations nor 
trends as a function of radial velocity, confirming that CoRoT-9b is a well-secured planet.

The 28 HARPS measurements we present here have an average precision of $9\ms$
and cover a time span of almost five years. 
This is a significant improvement compared with the 14 measurements gathered by D10
in a one-year time span. The 28 observations presented in Table~\ref{table_rv} 
include those of D10, but with slightly different numerical values as our data reduction is 
not exactly identical.

\begin{table}[t] 
\caption{HARPS radial-velocity measurements of CoRoT-9.}
\begin{tabular}{lccrcc}
\hline
BJD$_{\rm UTC}$ & RV & $\pm$$1\,\sigma$ & bisect.$^\dagger$ & Texp$^\star$ & S/N$^\star$$^\star$ \\
-2\,450\,000 & [$\kms$] & [$\kms$] & [$\kms$]  & [sec] \\
\hline
4730.5481 & 19.785 & 0.006 & -0.003 & 2700 &  18.3 \\       
4734.5653 & 19.797 & 0.008 & -0.008 & 2700 &  15.7 \\     
4739.5092 & 19.790 & 0.008 &  0.022 & 2700 &  16.1 \\     
4754.5002$\dagger$$\dagger$ & 19.811 & 0.011 &  0.021 & 3600 &  19.1 \\   
4766.5131 & 19.838 & 0.008 & -0.001 & 3600 &  17.3 \\             
4771.5094 & 19.865 & 0.014 &  0.036 & 3100 &  10.6 \\     
4987.7410$\dagger$$\dagger$ & 19.802 & 0.012 &  0.021 & 3600 &  16.0 \\   
4993.8179$\dagger$$\dagger$ & 19.787 & 0.007 &  0.000 & 3600 &  22.8 \\   
5021.6567$\dagger$$\dagger$ & 19.798 & 0.009 &  0.028 & 3600 &  19.7 \\   
5048.6495$\dagger$$\dagger$ & 19.850 & 0.007 &  0.030 & 3600 &  21.8 \\   
5063.5679 & 19.856 & 0.011 & -0.006 & 3600 &  16.5 \\     
5069.5573 & 19.849 & 0.006 &  0.008 & 3300 &  18.7 \\     
5077.5704$\dagger$$\dagger$ & 19.810 & 0.012 & -0.002 & 3000 &  16.0 \\   
5095.5408 & 19.789 & 0.006 &  0.011 & 3600 &  19.7 \\     
5341.9076 & 19.856 & 0.005 &  0.002 & 3600 &  23.3 \\     
5376.7009$\dagger$$\dagger$ & 19.787 & 0.014 & -0.016 & 3600 &  16.5 \\   
5400.6811$\dagger$$\dagger$ & 19.787 & 0.009 &  0.030 & 3600 &  20.3 \\   
5413.6660 & 19.800 & 0.005 &  0.006 & 3600 &  24.0 \\     
5423.6248 & 19.833 & 0.009 & -0.026 & 3600 &  15.9 \\     
5428.5913$\dagger$$\dagger$ & 19.839 & 0.015 & -0.048 & 3600 &  13.7 \\   
5658.8885 & 19.783 & 0.011 & -0.017 & 3600 &  13.5 \\     
5682.9153 & 19.779 & 0.006 &  0.022 & 2400 &  20.9 \\     
5713.8266 & 19.855 & 0.009 &  0.026 & 3600 &  15.7 \\     
5769.5810$\dagger$$\dagger$ & 19.794 & 0.009 & -0.005 & 3600 &  14.9 \\   
5824.5177 & 19.862 & 0.007 &  0.004 & 3600 &  19.4 \\     
6120.6337 & 19.838 & 0.011 &  0.003 & 3600 &  12.8 \\     
6469.6810 & 19.807 & 0.016 &  0.009 & 2700 &  13.9 \\     
6515.6036 & 19.802 & 0.008 &  0.006 & 3600 &  17.2 \\     
\hline
\multicolumn{6}{l}{$\dagger$: bisector spans; error bars are twice those of the RVs.} \\ 
\multicolumn{6}{l}{$\star$: duration of each individual exposure.} \\
\multicolumn{6}{l}{$\star\star$: signal-to-noise ratio per pixel at 550\,nm.} \\
\multicolumn{6}{l}{$\dagger$$\dagger$: measurements corrected for moonlight pollution.} \\ 
\label{table_rv}
\end{tabular}
\end{table}

\section{Bayesian data analysis}
\label{data_analysis}
To derive the system parameters, we carried out 
a Bayesian combined analysis of the space-based photometric data and 
ground-based HARPS RVs, using a differential evolution Markov chain Monte Carlo (DE-MCMC) 
technique \citep{Braak2006, 2013PASP..125...83E}. 
For this purpose, the epochs of the photometric and spectroscopic data were 
converted into the same unit $\rm BJD_{TDB}$ \citep{2010PASP..122..935E};  
we used Eq.~(4) in \citet{2014ApJ...788...92S} to perform this correction for 
the Spitzer data. Given the relatively large semimajor axis of CoRoT-9b, 
light travel time between the transit measurements and the stellar-centric frame 
(to which RV epochs are referred) amounts to $\sim3$~min and was taken into account to 
have all the data in the same reference frame (the system barycentre).

Our model consists of i) the Keplerian orbit to fit the RVs of the host star and ii) the CoRoT-9b transit model, 
for which we used the formalism of \citet{2002ApJ...580L.171M}.
The free parameters are as follows:
the transit epoch $T_{\rm 0}$; the orbital period $P$; 
the systemic radial velocity $V_{\rm r}$; the radial-velocity semi-amplitude $K$; 
$\sqrt{e}~{\cos{\omega}}$ and $\sqrt{e}~{\sin{\omega}}$
(e.g. \citealt{2011ApJ...726L..19A});
a RV jitter term $s_{\rm j}$ added in quadrature to the formal error bars 
to account for possible extra noise in the RV measurements; 
the transit duration from first to fourth contact $T_{\rm 14}$;
the ratios of the planetary-to-stellar radii $R_{\rm p}/R_{*}$ for both the CoRoT and Spitzer bandpasses;
the inclination $i$ between the orbital plane and the plane of the sky;
the two limb-darkening coefficients (LDC)
$q_{1}=(u_{a}+u_{b})^2$ and $q_{2}=0.5 u_{a} / (u_{a}+u_{b})$ \citep{2013MNRAS.435.2152K}, 
where $u_{\rm a}$ and $u_{\rm b}$ 
are the coefficients of the limb-darkening quadratic law\footnote{
$I(\mu)/I(1)=1-u_{\rm a}(1-\mu)-u_{\rm b}(1-\mu)^2$, where $I(1)$ is the 
specific intensity at the centre of the disk and $\mu=\cos{\gamma}$, 
$\gamma$ being the angle between the surface normal and the line of sight.}, for both the CoRoT and Spitzer bandpasses;
and two contamination factors, one for the CoRoT transit observed in 2011 in imagette mode (see Sect.~\ref{photometric_data}) 
and the other for the Spitzer transit observed in 2010 to account for the significant difference in 
depth with respect to the 2011 Spitzer transit (Sect.~\ref{photometric_data}).
Uniform priors were set on all parameters, in particular with bounds of [0, 1] for  
$q_{1}$ and $q_{2}$ \citep{2013MNRAS.435.2152K}, with a lower limit of zero for 
$K$ and $s_{\rm j}$, and an upper bound of 1 for the eccentricity; the lower limit 
of zero simply comes from the choice of fitting $\sqrt{e}~{\cos{\omega}}$ and 
$\sqrt{e}~{\sin{\omega}}$). 
Transit fitting was also performed for each individual transit to compute transit timing variations 
(Sect.~\ref{CoRoT-9b_alone}) by fixing $e$ and $\omega$ to the values found 
with the combined analysis and by imposing a Gaussian prior on transit duration for the 
partial CoRoT transit.

The DE-MCMC posterior distributions of the model parameters 
were determined by 
i) maximising a Gaussian likelihood function (see e.g. Eqs.~9 and 10 in \citealt{2005ApJ...631.1198G}); 
ii) adopting the Metropolis-Hastings algorithm to accept or reject the proposed steps;   
and iii) following the prescriptions given by \citet{2013PASP..125...83E} for the employed number of chains (twice the number of free parameters), 
the removal of burn-in steps, and the criteria for convergence and proper mixing of the chains.
As usual, the medians and the $15.86\%$ and $84.14\%$ quantiles of the posterior distributions  
are taken as the best values and $1\sigma$ uncertainties of the 
fitted and derived parameters. For parameters consistent with zero, we provide the $1\sigma$ 
upper limits computed as the $68.27\%$ confidence intervals starting from zero. 

The stellar density as derived from the transit fitting, along with the effective temperature and metallicity of CoRoT-9
that are reported in D10, were interpolated to the Yonsei-Yale evolutionary tracks \citep{2004ApJS..155..667D} to find 
the most likely stellar, hence planetary, parameters and their associated uncertainties 
\citep[e.g.][]{2007ApJ...664.1190S, 2014A&A...572A...2B}.

\section{Results} 
\subsection{System parameters.} 
\label{system_parameters} 
\begin{figure}[t]
\vspace{-0.3cm}
\hspace{0.7cm}
\begin{minipage}{9cm}
\includegraphics[width=6cm, angle=90]{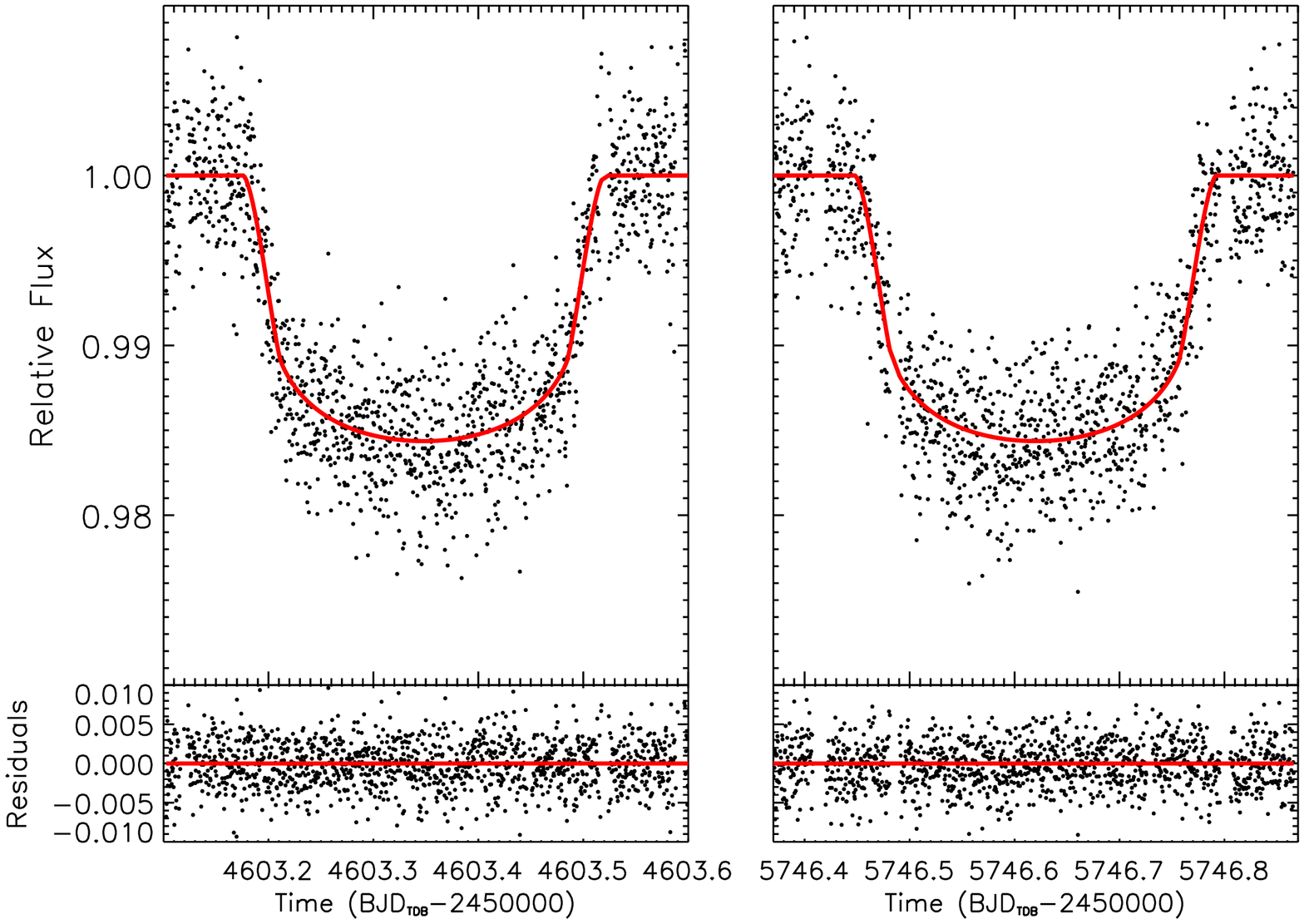}
\includegraphics[width=6cm, angle=90]{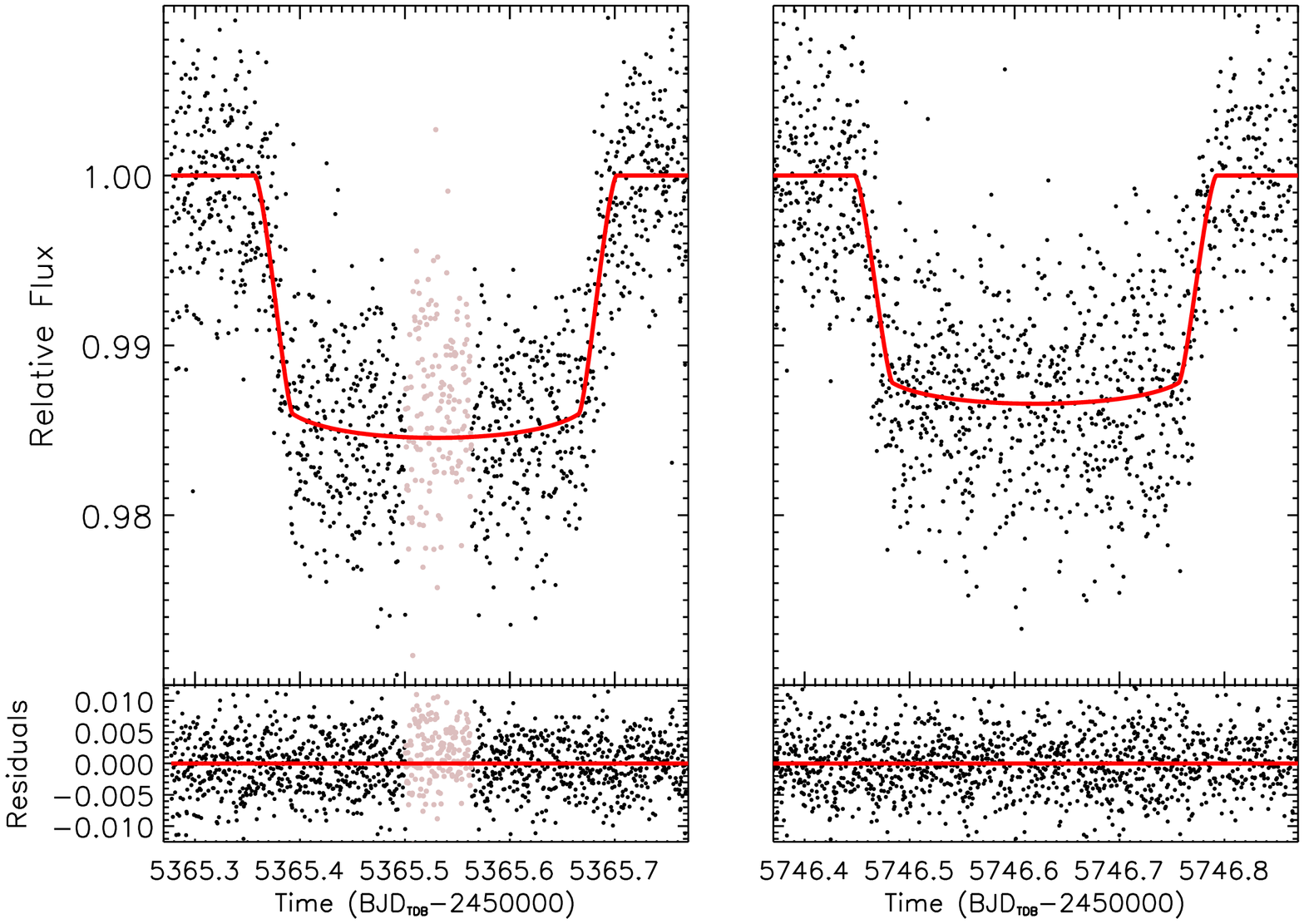}
\end{minipage}
\hspace{-0.5cm}
\caption{
\emph{Top panel}: CoRoT-9b full transits as observed by the CoRoT satellite with a 32~s cadence (optical band) and 
the best-fit model (red solid line).
\emph{Bottom panel}: The other two transits observed with Spitzer at $4.5~\rm \mu m$ with a 31~s cadence along with the transit model (red solid line). We note the different transit depths and the ``bump'' at the middle of the first transit (grey points), which we attribute to uncorrected systematics affecting the 2010 Spitzer transit (see text). }
\label{fig_transits}
\end{figure}

\begin{figure*}[t]
\centering
\begin{minipage}{16cm}
\includegraphics[width=6cm, angle=90]{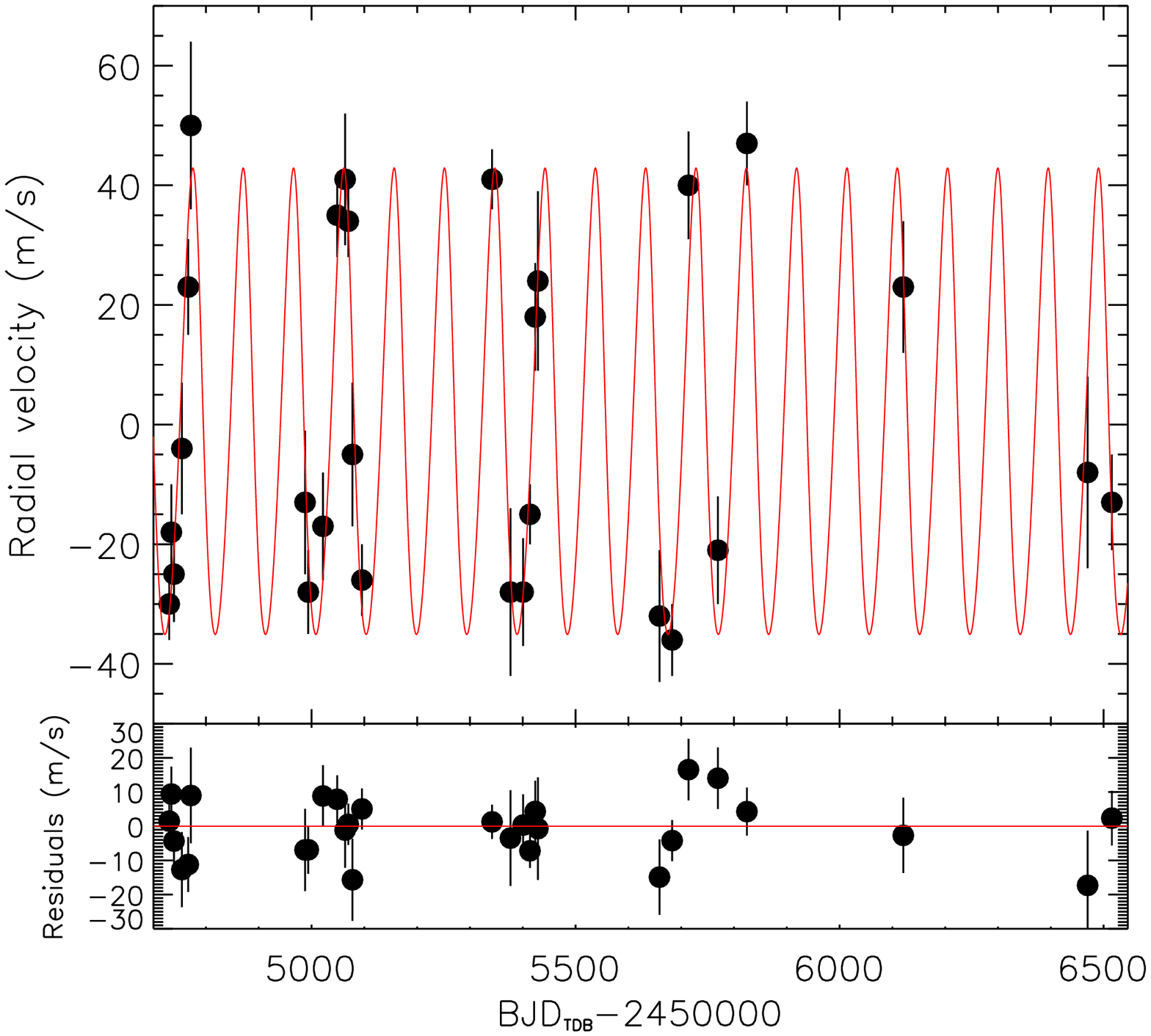}
\includegraphics[width=6cm, angle=90]{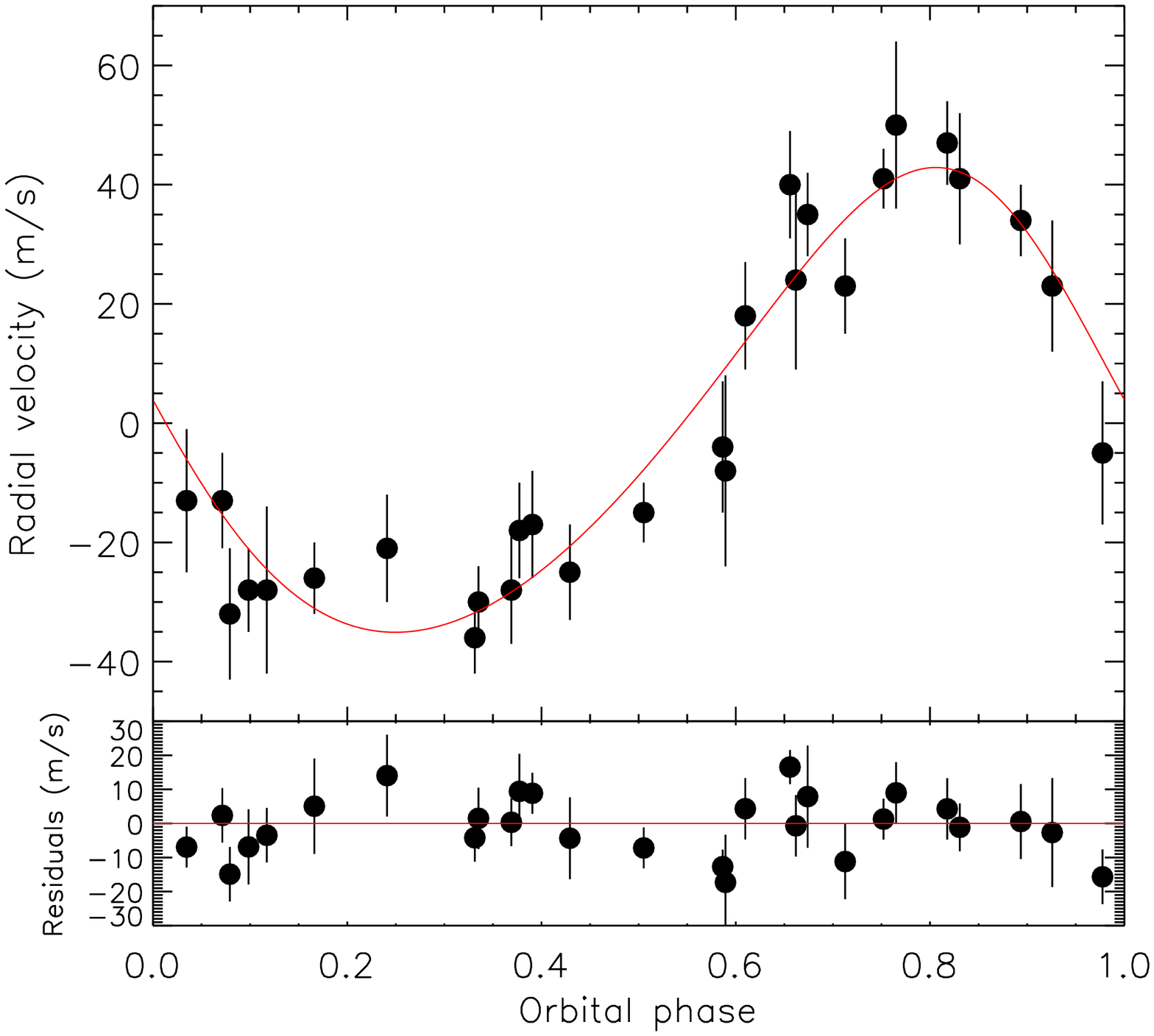}
\vspace{0.3cm}
\caption{
\emph{Left panel}: HARPS RVs of CoRoT-9 as a function of time and the best Keplerian model (red solid line).
\emph{Right panel}: The same as the left panel but as a function of the orbital phase (transits occur at phase equal to zero/one). }
\label{fig_RV}
\end{minipage}
\end{figure*}

Fitted and derived system parameters and their $1\sigma$ error bars 
are given in Table~\ref{starplanet_param_table}. 
They are fully consistent, that is within $1\sigma$, with those reported by D10. 
Figures~\ref{fig_transits} and \ref{fig_RV} show the best-fit models 
to the full CoRoT and Spitzer transits and the RV data, respectively. 

The values of $R_{\rm p}/R_{*}$ in the CoRoT and Spitzer bandpasses agree within $1\sigma$, 
the former being slightly higher. Flux contamination by background stars in the CoRoT imagette mode
was found to be negligible, that is $< 0.4\%$; the dilution factor for the first Spitzer transit 
to account for its larger depth (Sect.~\ref{photometric_data}) is $13.2\pm1.5\%$. 
The fitted limb-darkening coefficients agree well with the theoretical values computed by 
\citet{2011A&A...529A..75C} both for the CoRoT and Spitzer bandpasses.

One of the most remarkable results of our analysis is that, by doubling 
the number of collected HARPS RVs, the significance of the small eccentricity of CoRoT-9b
increases from 2.75 (D10) to 3.6$\sigma$, where $e=0.133^{+0.042}_{-0.037}$. 
To investigate whether the CoRoT-9b small eccentricity is bona fide or might be spurious, 
we used a Bayesian model comparison to compute the relative probabilities of the eccentric versus circular orbit models 
by fitting a Keplerian model to the HARPS RVs with Gaussian priors imposed on the photometric 
transit time and orbital period (Table~\ref{starplanet_param_table}). 
For this purpose, we computed the Bayesian evidence for both the circular and eccentric models 
with the \citet{Perrakis201454} method and its implementation as described in \citet{2016A&A...585A.134D}.
We found $B_{\rm ecc, circ}=124 \pm 8$ in favour of the eccentric model, which means that the eccentric model is 
$\sim124$ times more likely than the circular model. According to \citet{doi:10.1080/01621459.1995.10476572}, 
this value of Bayes factor largely exceeds the threshold ($B=20$) of strong 
evidence for a more complex model, in our case for the eccentric model with respect to the circular model. 
This is a significant improvement with respect
to the old HARPS data published by D10 because those data only yield $B_{\rm ecc, circ}=4.3 \pm 0.4$, which 
is below the aforementioned threshold to claim a significant eccentricity.

\begin{table*}[t]
\centering
\caption{CoRoT-9 system parameters.}            
\begin{minipage}[t]{13.0cm} 
\renewcommand{\footnoterule}{}                          
\begin{tabular}{l l }        
\hline\hline                 
\emph{Stellar parameters}  &  \\
\hline
Stellar mass [\Msun] &  $ 0.96 \pm 0.04 $  \\
Stellar radius [\Rsun] & $ 0.96 \pm 0.06$   \\
Stellar density $\rho_{*}$ [$ \rm g\;cm^{-3}$] & $1.51 \pm 0.30$ \\
Age $t$ [Gyr]  & $6 \pm 3$  \\
Effective temperature $T_{\rm{eff}}$[K]~$^a$ & 5625 $\pm$ 80 \\
Stellar surface gravity log\,$g$ [cgs]~$^a$ &  4.54  $\pm$ 0.09  \\
Stellar metallicity $[\rm{Fe/H}]$ [dex]~$^a$ & -0.01  $\pm$ 0.06 \\
Stellar rotational velocity $V \sin{i_{*}}$ [\kms]~$^a$ & $< 3.5$  \\
CoRoT limb-darkening coefficient $q_{1}$  &  $0.40^{+0.13}_{-0.10}$ \\
CoRoT limb-darkening coefficient $q_{2}$  &  $0.33^{+0.13}_{-0.10}$  \\
CoRoT limb-darkening coefficient $u_{a}$  &  $0.41 \pm 0.09$ \\
CoRoT limb-darkening coefficient $u_{b}$  &  $0.22 \pm 0.17$  \\
Spitzer limb-darkening coefficient $q_{1}$  &  $0.035^{+0.037}_{-0.020}$ \\
Spitzer limb-darkening coefficient $q_{2}$  &  $0.39^{+0.36}_{-0.27}$  \\
Spitzer limb-darkening coefficient $u_{a}$  &  $0.14^{+0.10}_{-0.08}$ \\
Spitzer limb-darkening coefficient $u_{b}$  &  $0.037^{+0.14}_{-0.11}$  \\
Systemic velocity  $V_{\rm r}$ [\kms] & $19.8150 \pm 0.0017$ \\
Radial-velocity jitter $s_{\rm j}$ [\ms] & $<3.9$ \\
& \\
\hline
\emph{Transit and orbital parameters}  &  \\
\hline
Orbital period $P$ [d] & $95.272656 \pm 0.000068$ \\
Transit epoch $T_{ \rm 0} [\rm BJD_{TDB}-2450000$]~$^b$ & 5365.52723 $\pm$ 0.00037  \\
Transit duration $T_{\rm 14}$ [d] & $0.3445^{+0.0021}_{-0.0018}$  \\
CoRoT bandpass radius ratio $R_{\rm p}/R_{*}$~$^c$ & $0.11402_{-0.00085}^{+0.00095}$   \\
Spitzer bandpass radius ratio $R_{\rm p}/R_{*}$~$^c$ & $0.11284_{-0.00092}^{+0.00086}$   \\
Inclination $i$ [deg] & $89.900_{-0.084}^{+0.066}$  \\
$a/R_{*}$ & $89.9 \pm 5.9$  \\
Impact parameter $b$ & $0.16^{+0.11}_{-0.09}$  \\
$\sqrt{e}~\cos{\omega}$ &  $0.29_{-0.08}^{+0.06} $ \\
$\sqrt{e}~\sin{\omega}$  &  $0.19_{-0.16}^{+0.12} $ \\
Orbital eccentricity $e$  &  $0.133^{+0.042}_{-0.037}$   \\
Argument of periastron $\omega$ [deg] & $41^{+31}_{-24}$ \\
Radial-velocity semi-amplitude $K$ [\ms] & $39.0 \pm 2.4$ \\
& \\
\hline
\multicolumn{2}{l}{\emph{Planetary parameters}} \\
\hline
Planet mass $M_{\rm p} ~[\Mjup]$  &  $0.84 \pm 0.05$  \\
Planet radius $R_{\rm p} ~[\Rjup]$  &  $1.066^{+0.075}_{-0.063}$  \\
Planet density $\rho_{\rm p}$ [$\rm g\;cm^{-3}$] &  $0.86^{+0.18}_{-0.16}$  \\
Planet surface gravity log\,$g_{\rm p }$ [cgs] &  $3.26 \pm 0.06$  \\
Orbital semimajor axis $a$ [AU] & $0.4021 \pm 0.0054$   \\
Equilibrium temperature $T_{\rm eq}$ [K]~$^d$  & $420 \pm 16$ \\
\hline       
\hline
\vspace{-1.3cm}
\footnotetext[1]{\scriptsize values from D10. } \\
\footnotetext[2]{\scriptsize in the planet-reference frame.} \\
\footnotetext[3]{\scriptsize two dilution factors were fitted along with the radius ratios for the transits observed by Spitzer in 2010 and 
CoRoT in 2011 (see text for more details); their values were found to be $13.2 \pm 1.5 \%$ (Spitzer) and $<0.4\%$ (CoRoT).} \\
\footnotetext[4]{\scriptsize black-body equilibrium temperature assuming a null Bond albedo and uniform heat redistribution to the nightside.} \\
\end{tabular}
\end{minipage}
\label{starplanet_param_table}  
\end{table*}

\subsection{Is CoRoT-9b alone?}
\label{CoRoT-9b_alone}
\begin{figure}[h]
\centering
\includegraphics[width=6cm, angle=90]{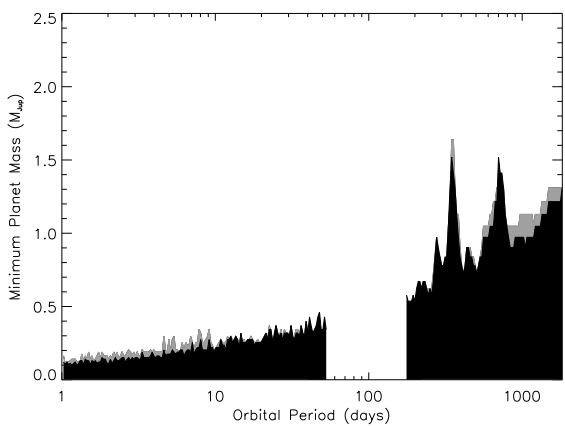}
\caption{
Upper limits on the minimum mass of a possible planetary companion of CoRoT-9b with $99\%$ confidence level 
as a function of orbital period up to 5 years. Black and shaded areas refer to circular and eccentric orbits
of the hypothetical companion; the maximum allowed eccentricity for each period was derived from dynamical stability criteria 
(see text for more details). The empty region around the orbital period of CoRoT-9b ($P=95.27$~d) 
is dynamically unstable for the minimum planetary masses detectable with our HARPS RV data. 
}
\label{fig_det_limits}
\end{figure}

As shown in Fig.~\ref{fig_RV} (lower panels), the residuals of the RVs appear flat, 
not showing any significant variation attributable 
to the presence of an additional planetary companion. From these residuals spanning almost five years, we
computed the detection limits given the sampling and precision of the HARPS RVs.
To this end, we injected into the residuals artificial Keplerian signals of a hypothetical companion by varying 
its minimum mass and orbital period with logarithmic grids from $1~\Mearth$ to $5~\Mjup$ and from 1~d to 5~yr, respectively.
For a given minimum mass and orbital period, we generated 500 different realisations of Keplerian models with randomly chosen values of 
periastron time, argument of periastron, and eccentricity; the maximum allowed eccentricity for each combination of mass and period of the 
simulated planet was determined from the semi-empirical stability criteria of \citet{2013MNRAS.436.3547G} in the most
conservative case, that is by assuming that the orbit of CoRoT-9b and its hypothetical companion are anti-aligned. 
We carried out simulations by considering only a circular orbit for the second planet as well. 
Then we made use of both the F-test and the $\chi^{2}$ statistics to exclude 
planetary companions of a given mass and period that would induce RV variations that are incompatible 
at $99\%$ confidence level with the observed RV residuals (Fig.~\ref{fig_RV}).
In such a way, we derived the upper limits on the minimum mass 
of a putative second planet as a function of orbital period. 
These are shown in Fig.~\ref{fig_det_limits} for circular 
(black area) and eccentric (grey contours) orbits. 
The region around the orbital period of CoRoT-9b is empty because it is dynamically unstable 
for the planetary minimum masses that are detectable with the gathered HARPS RVs.
The peaks at $P\sim1$ and $2$~yr are due to the temporal sampling of the RV measurements 
that is inevitably affected by the object's visibility. 
From these detection limits, we are able to rule out the presence of massive companions 
of CoRoT-9b, specifically companions with $\mp \sin{i}\gtrsim$0.25, 1.2, and 1.4~$\Mjup$ 
and $P\sim$10~d, 3~yr, and 5~yr, respectively.

\begin{figure}[h]
\centering
\includegraphics[width=6cm, angle=90]{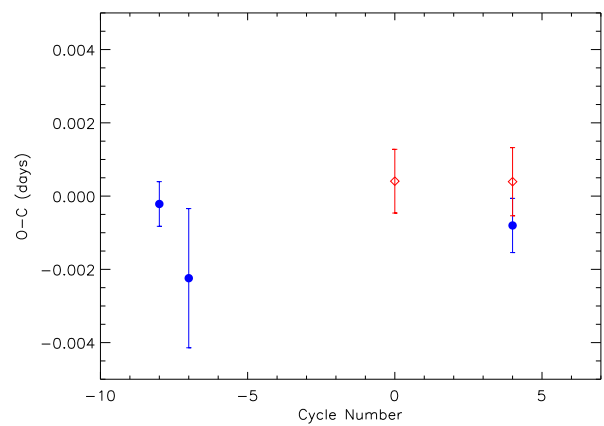}
\caption{
Residuals of the mid-transit epochs of CoRoT-9b vs. the linear ephemeris reported in Table~\ref{starplanet_param_table}. 
Blue filled circles refer to the transits observed by CoRoT; empty red diamonds indicate the two Spitzer transits. 
The value with the largest error bar comes from a partial CoRoT transit. 
}
\label{fig_TTV}
\end{figure}

The TTVs show no significant variations from a linear ephemeris either (see Fig.~\ref{fig_TTV} and Table~\ref{table_TTV}).
This also suggests the absence of a strong perturber, even though only four transit epochs 
could be determined.

By considering the r.m.s. of the CoRoT light curve (Sect.~\ref{photometric_data}) 
and $S/N=10$ as the transit detection threshold \citep{2012A&A...547A.110B}, 
we can exclude the presence of inner (coplanar) transiting planets in the system with 
$\rp \gtrsim 1.3$, 2.0, 2.3, and 2.6~$\Rearth$ and orbital periods $P=1$, 10, 25, and 50~d, respectively.

\begin{table}[h]
\caption{Times of CoRoT-9b mid-transits.}            
\renewcommand{\footnoterule}{}                          
\begin{tabular}{l l l}      
\hline \hline
Time & Uncertainty & Instrument \\
$[\rm BJD_{TDB}-2450000]$ & [days] & \\
\hline
4603.34577      &       0.00061         & CoRoT         \\
4698.6164       &       0.0019          & CoRoT \\
5746.61705      &       0.00074         & CoRoT \\
5365.52764      &       0.00087         & Spitzer \\
5746.61825      &       0.00093         & Spitzer \\
\hline \hline
\end{tabular}
\label{table_TTV}
\end{table}

\section{Origin of the non-circular orbit of  Corot-9b }
\label{dynamical_simulations}
We now consider the question of the evolution of CoRoT-9b.  Given its clearly detected non-zero eccentricity, we place constraints on the dynamical history of the CoRoT-9 system.  We first discuss a broad range of potential formation models, culminating with what we consider to be the most likely candidate: planet-planet scattering.  Next we present a suite of scattering simulations to reproduce the orbit of CoRoT-9b.  Finally, we construct a plausible formation scenario for the system.  Our results provide motivation to measure the sky-projected obliquity of the CoRoT-9 system.

\subsection{Possible evolutionary scenarios for the CoRoT-9 system}
\label{evolutionary_scenarios}
There exist a number of mechanisms that could explain the origin of the CoRoT-9 system.

{\bf 1. In situ formation of CoRoT-9b from local material at $\sim$0.4 AU.}  It is possible that there was enough material in the disk for the planet to grow a core of several Earth masses and to accrete gas from the disk~\citep[e.g.][]{ikoma01,hubickyj05,raymond08,batygin16}.  However, if the planet formed in situ, it is hard to understand why only one planet should have formed.  And even if there exist additional planets that are too small to detect, why would CoRoT-9b have an eccentric orbit?  Additional low-mass planets cannot pump the eccentricity of CoRoT-9b to its observed value.  Isolated in situ accretion is implausible.

{\bf 2. Formation of CoRoT-9b farther from the star followed by gas-driven inward migration.}  Migration is indeed a likely -- and unavoidable -- consequence of planet-disk interaction~\citep{goldreich80,ward86,lin86,papaloizou06}.  Migration is usually directed inward and can plausibly explain the origin of close-in planets of a wide range of masses~\citep{kley12,baruteau14}.  However, simulations show that orbital migration of a single planet universally lowers the orbital eccentricity of a planet~\citep{tanaka04,cresswell08,bitsch10} except in the extreme case of a very massive planet ($\mp\sim5-10~\Mjup$) in a very massive disk~\citep{papaloizou01,kley06,dunhill13}.  The  non-zero eccentricity of CoRoT-9b appears to rule out a solitary migration scenario. 

{\bf 3. Inward migration of CoRoT-9b driven by secular forcing from a more distant giant planet.}  \cite{petrovich16} proposed that WJs are driven inward by a combination of secular forcing from a distant companion and tidal dissipation~\citep[see also][]{dawson14}.  In this model, the eccentricity of the inner planet is periodically driven to such high values -- and its perihelion distance to such low values -- that tidal dissipation shrinks the orbit of the planet.  The inner planet is thus driven inward in periodic bursts.  However, this model requires the presence of a second planet on a more distant, very eccentric orbit.  There is no hint of such a distant perturber (see Sect.~\ref{CoRoT-9b_alone}). While constraints on additional planets in the CoRoT-9 system cannot completely rule out this model, it is worth noting that the model struggles to produce WJs with modest ($e \lesssim 0.2$) eccentricities.  This scenario appears unlikely to explain the CoRoT-9 system.

{\bf 4. CoRoT-9b as the survivor of a dynamical instability.}  The planet-planet scattering model can explain the broad eccentricity distribution of observed giant exoplanets~\citep[e.g.][]{adams03,chatterjee08,juric08,ford08,raymond10}.  This model proposes that the observed planets are the survivors of violent dynamical instabilities in which multiple planets underwent close gravitational encounters.  During these gravitational scattering events, one or more planets are lost, usually by ejection into interstellar space~\citep[although their numbers are too low to explain the abundance of free-floating gas giants;][]{veras12}. The CoRoT-9 system could have formed with one or more additional planets whose orbits became unstable.  Given the proximity of CoRoT-9b to the star, the planets may have migrated inward and then become unstable when the gas disk dissipated~\citep[see e.g.][]{ogihara09,cossou14}.  

The planet-planet scattering mechanism operates when the gravitational kick of a planet is strong enough that scattering dominates over accretion.  This is often quantified with the so-called Safronov number $\Theta$ \citep{1969Icar...10..109S}, which is defined as the ratio of the escape speed from the surface of a planet to the escape speed from the star, or 
\begin{equation}
\Theta^2 = \left(\frac{G \mp}{\rp}\right) \left(\frac{a}{G M_\star}\right) = \left(\frac{\mp}{M_\star}\right) \left(\frac{a}{\rp}\right),
\end{equation}
where $\mp$ and $M_\star$ are the planetary and stellar masses, respectively, $\rp$ is the planet's radius and $a$ is the orbital radius.  Giant exoplanets with larger $\Theta$ are observed to have higher eccentricities~\citep{ford01,ford08}.  
For the case of CoRoT-9b, $\Theta=0.66$, that is, close to unity. 
This puts the planet right at the boundary between the scattering and accretionary regimes. While a scattering origin appears plausible to explain the origin of the non-zero eccentricity of CoRoT-9b, it is worth checking with numerical simulations.

\subsection{Scattering simulations to explain the orbit of CoRoT-9b}
\label{scattering_simulations}
We ran a suite of numerical simulations of planet-planet scattering.  The goal of these simulations was to test whether the orbit of CoRoT-9b can be reproduced by planet scattering.  In this context, CoRoT-9b would represent the survivor of a dynamical instability that removed another planet from the system, likely by dynamical ejection.  For simplicity we only included one additional planet.

Each simulation contained a star with the properties of CoRoT-9 orbited by two planets: a CoRoT-9b analogue with the actual mass of the planet, and a second planet. CoRoT-9b analogues were initially placed at 0.46~AU on circular orbits. 
This orbital radius was chosen so that, after ejecting a 50~$M_\oplus$ companion, CoRoT-9b would end up on its actual orbit. This initial radius would shift by up to $\sim 0.1$~AU for CoRoT-9b to end up on its actual orbit for the range of extra planet masses considered. While a different initial radius for CoRoT-9b would modestly change its starting Safronov number, this would not affect the outcome of scattering. In fact, we show below that, while changing the Safronov number of CoRoT-9b impacts the branching ratios of different outcomes (i.e. a lower $\Theta^2$ implies a higher rate of collisions), it does not change the outcomes themselves (i.e. the final eccentricity of CoRoT-9b is not sensitive to the value of $\Theta^2$). 
The extra planet was placed on an exterior orbit within 2\% of the 3:4 or 4:5 mean motion resonances.\footnote{
For one set of simulations we also tested placing the extra planet on a near-resonant orbit {\em interior} to the CoRoT-9b analogues and found no measurable difference in outcome.} The reason for this was to remain consistent with a migration origin for the planets, which generally implies resonant capture~\citep[e.g.][]{kley12}, while being dynamically unstable~\citep{marchal82,gladman93}. The orbits of the planets were given a randomly chosen small ($<1^\circ$) initial mutual inclination.  

We tested two parameters: the mass of the second planet and the physical density of the planets.  We considered extra planet masses $M_{\rm extra}=25$, 50, 75, 100, 200 and 267~$\Mearth$ (the mass of CoRoT-9b is 267~$\Mearth$).  For an extra planet mass of $50~\Mearth$ we also tested physical densities for the planets of 0.5, 1.0, and 2.0 $\rm g\, cm^{-3}$.  
In all other simulations the densities of both planets were fixed at 1.0 $\rm g\, cm^{-3}$; even though this value is consistent with the measured density of CoRoT-9b within $1\sigma$ ($0.86^{+0.18}_{-0.16}~\rm g\,cm^{-3}$), we show below that the density has no effect on the outcome.  For each set we ran 100 simulations.

Each simulation was integrated for 10 million years or until the system became unstable and one planet was removed.  We used the Mercury hybrid integrator~\citep{chambers99} with a timestep of 0.1 days.  This timestep was small enough to accurately resolve orbits that collided with the star, which were assumed in the calculations to have radii of 0.01 AU \citep[see Appendix A of][for representative numerical tests]{raymond11}.  Planets were removed from the system if their orbital radii reached 1000 AU.  When this happened, collisions between the planets were treated as inelastic mergers.

\begin{figure}
\centering
\includegraphics[width=9cm]{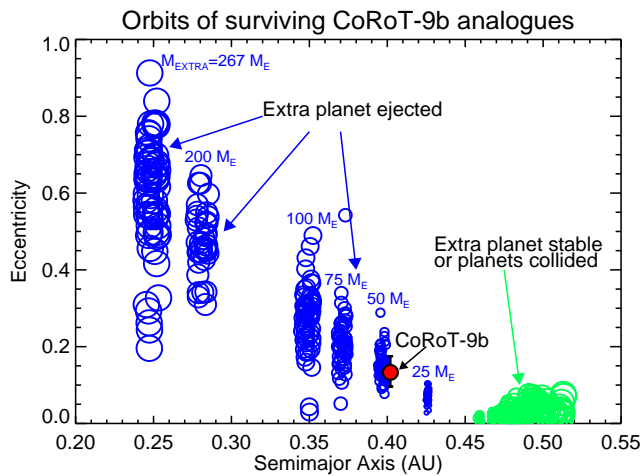}
\caption{Orbital semimajor axes and eccentricities of surviving CoRoT-9b analogues from our numerical simulations.  The blue planets are those that remained after the second planet was ejected; each vertical band corresponds to a specific mass for the extra planet.  The green circles are simulations in which either there was no instability or the two planets collided.  The measured orbit of CoRoT-9b is shown with the red symbol; the error bars are $1\sigma$. }
\label{fig:ae}
\end{figure}

Figure~\ref{fig:ae} shows the orbits of CoRoT-9b analogues at the end of the simulations.  The small clump at 0.46 AU represents systems that remained stable; in those cases CoRoT-9b remained on a circular orbit.  In simulations in which the two planets collided, CoRoT-9b analogues are at slightly larger orbital radii (0.47-0.52 AU) and with small eccentricities (typically $e \lesssim 0.05$).  A collision between CoRoT-9b and an extra planet is clearly inconsistent with the measured eccentricity of CoRoT-9b.

When the orbits of the planets become unstable and the extra planet is ejected, CoRoT-9b analogues are shifted inward from their initial orbits.  The outcomes are radially segregated by the mass of the extra planet because of the mass dependence of the orbital energy exchanged when the extra planet was ejected.  In simple terms, CoRoT-9b feels a mass-dependent recoil from ejecting the extra planet.  Thus, each vertical ``stripe'' in Fig.~\ref{fig:ae} represents the outcome of a set of simulations with a specified extra planet mass.  It is important to note that these simulations were designed to reproduce the eccentricity of CoRoT-9b, not its semimajor axis.  Thus, even though the simulations with $M_{\rm extra} = 50~\Mearth$ are the closest match in semimajor axis, other sets of simulations could easily match the semimajor axis of CoRoT-9b; for example, 
the simulations with $M_{\rm extra} = 75~\Mearth$ would match if CoRoT-9b had started at $\sim$0.49 AU instead of 0.46 AU.

The eccentricities of surviving CoRoT-9b analogues correlate with the mass of the ejected planet.  The ejection of a 25~$\Mearth$ planet only excites CoRoT-9b analogues to an eccentricity of $\sim0.05,$ whereas ejecting a planet comparable in mass can excite the eccentricities of the planets up to 0.9.  Nonetheless, there is a broad range in eccentricities of CoRoT-9b analogues for each set of simulations with a specified extra planet mass.  This is because the final eccentricity depends on the details of the last close encounters between the planets, whose alignment is stochastic in nature.

\begin{figure}
\centering
\includegraphics[width=8cm]{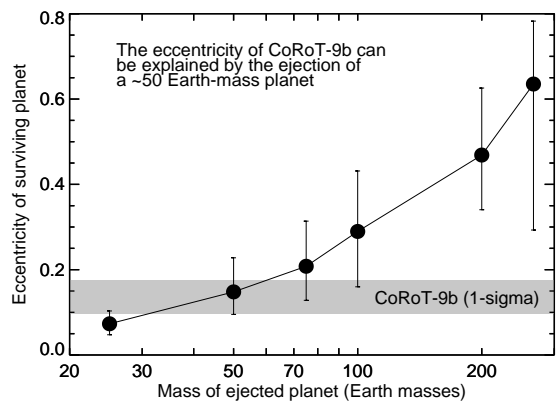}
\caption{Eccentricity of surviving planets as a function of the mass of the ejected planet.  The large points show the median for each set and the error bars show the 5th-95th percentile of outcomes. The shaded area shows the $\pm1-\sigma$ range of the measured eccentricity of CoRoT-9b.}
\label{fig:e-mextra}
\end{figure}

Figure~\ref{fig:e-mextra} shows the eccentricity of CoRoT-9b analogues after ejecting an extra planet.  It is clear that the set of simulations with $M_{\rm extra} = 50~\Mearth$ most readily produces CoRoT-9b analogues with the measured eccentricity.  However, there is a small tail of outcomes with $M_{\rm extra} = 25~\Mearth$ that overlaps with the lower allowed values for CoRoT-9b.  At larger masses, a significant fraction of simulations with $M_{\rm extra} = 75~\Mearth$ overlap with the higher allowed values, and even some simulations with $M_{\rm extra} = 100~\Mearth$ are allowed by observations.  

Planet-planet scattering also excites the planetary orbital inclinations \citep{juric08,chatterjee08,raymond10}.  Indeed, scattering of planets to extremely high eccentricities followed by tidal dissipation has been proposed as a mechanism to produce hot Jupiters whose orbits are misaligned with the stellar equator~\citep{nagasawa08,beauge12}. 

\begin{figure}
\centering
\includegraphics[width=9cm]{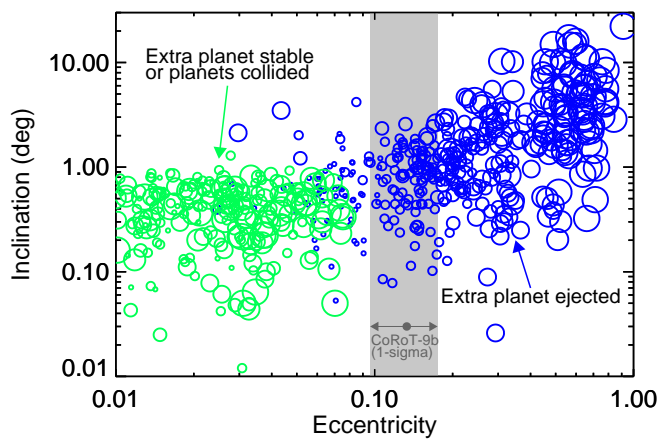}
\caption{Orbital inclination (zero~degrees correspond to the orbital plane prior to scattering phase) vs. eccentricity for CoRoT-9b analogues at the end of our simulations. As in Fig.~\ref{fig:ae}, blue circles represent planets that survived an instability whereas green circles represent planets that either remained on stable orbits or collided with the extra planet (recall that the initial inclination was randomly chosen between zero and $1^\circ$).  The size of the circle correlates with the mass of the extra planet.  The shaded region represents the $1\sigma$ allowed eccentricities for CoRoT-9b.}
\label{fig:ei}
\end{figure}

Figure~\ref{fig:ei} shows the inclinations of surviving CoRoT-9b analogues. None of the planets that match the eccentricity of CoRoT-9b have inclinations larger than $3^\circ$. Inclinations above $10^\circ$ were only produced in the most energetic scattering events, which also stranded CoRoT-9b analogues on orbits much more eccentric than the real planet.

Our simulations thus predict that the orbit of CoRoT-9b should be in the same plane as it started, to within a few degrees.  If we assume that plane to have been aligned with the stellar equator, then this naturally predicts that Rossiter-McLaughlin measurements should find a low stellar obliquity for CoRoT-9, i.e. an alignment between the planetary orbital plane and the stellar equator.

But what would it mean if Rossiter-McLaughlin measurements found a non-zero stellar obliquity?  Given the arguments presented above, we think that planet scattering is by far the most likely origin for the eccentric orbit of CoRoT-9b.  Assuming that scattering indeed took place, a measured non-zero stellar obliquity would imply that the planetary orbital plane was already misaligned {\it prior to the scattering phase}.  This would be strong indirect evidence for misalignment of the protoplanetary disk of CoRoT-9.  
Some candidate mechanisms to tilt disks are chaotic star formation \citep{2010MNRAS.401.1505B}, 
magnetic star-disk interactions \citep{2011MNRAS.412.2790L}, and perturbations from a distant stellar binary~\citep{batygin12,batygin13,lai14}. 
Concerning the latter, even temporary binarity during the embedded cluster phase may tilt the disk.

Finally, we also tested the effect of the planetary density $\rho$ on the outcome of scattering simulations.  This could be important because the density is linked to the radius of the planet and therefore its escape speed, and thus the Safronov number $\Theta$.

In the simulations presented above we assumed that both the CoRoT-9b analogues and the extra planet had densities of  $\rho = 1~\rm g\, cm^{-3}$.  We ran three sets of 100 simulations each with $M_{\rm extra} = 50~\Mearth$ and varying the density (of both planets) between 0.5 and 2.0~$\rm g\, cm^{-3}$.  This corresponded to a range in $\Theta^2$ between 0.55 and 0.87.  All other aspects of the simulations were as above.  

\begin{figure}
\centering
\includegraphics[width=9cm]{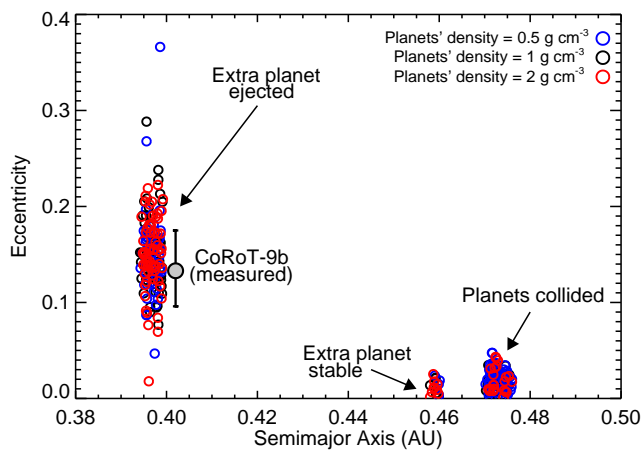}
\caption{Orbital eccentricity vs. semimajor axis for surviving CoRoT-9b analogues in three sets of simulations that varied the densities of the planets.  The blue, black, and red symbols represent simulations in which $\rho = $ 0.5, 1.0, and 2.0~$\rm g\, cm^{-3}$, respectively. }
\label{fig:dens}
\end{figure}

Figure~\ref{fig:dens} shows the surviving planets in the three sets of simulations with different densities.  At a glance the outcomes of the simulations appear similar.  Indeed, Kolmogorov-Smirnov tests show that, after ejecting the extra planet, the distribution of eccentricities of CoRoT-9b analogues in all three sets are consistent with having been drawn from the same distribution.

Yet the branching ratios between outcomes did depend on the planet density.  In the set of simulations with $\rho =  0.5~\rm g\, cm^{-3}$, collisions between the two planets were more than twice as common (53 collisions versus 24) as in the set of simulations with $\rho =  2~\rm g\, cm^{-3}$.  Ejections were significantly less common in the systems with low-density planets (42 ejections for $\rho =  0.5~\rm g\, cm^{-3}$ vs. 67 for  $\rho =  2.0 ~\rm g\, cm^{-3}$).  The simulations with  $\rho =  1.0~\rm g\, cm^{-3}$ were intermediate in both cases.

We can conclude from this simple numerical experiment that, while the planet density affects the probability of a given outcome, it has little effect on the details of that outcome.

\subsection{A plausible evolutionary history for CoRoT-9b}

Given the discussion above and the results of our scattering simulations, we propose the following evolutionary history for the CoRoT-9 system.
\begin{itemize}
\item CoRoT-9b formed relatively late in the lifetime of the protoplanetary disk, probably at a few AU, and migrated inward.  Its migration stopped at $\sim$0.5 AU because the disk was starting to photo-evaporate~\citep{alexander12}.
\item A second, $\sim50~\Mearth$ planet formed in the system.  This may have been a second core that formed farther out, migrated inward, and caught up with CoRoT-9b \citep[as in e.g. the Grand Tack model for the Solar System;][]{walsh11}, or perhaps a planet whose growth was accelerated by the migration of CoRoT-9b via the ``snowplow" effect of migrating resonances~\citep{zhou05,fogg05,raymond06,mandell07}.  The extra planet became locked in~\citep[or near;][]{adams08} mean motion resonance with CoRoT-9b.
\item The orbits of CoRoT-9b and the extra planet became unstable after the dispersal of the gaseous disk.  The two planets scattered off of each other, leading to the ejection of the smaller planet.  Nearby small planets were destroyed during the instability~\citep{raymond11,raymond12}. The measured eccentricity of CoRoT-9b is essentially a scar from this instability.
\end{itemize}

\section{Summary and conclusions}
Thanks to three new space-based observations of transits of CoRoT-9b with CoRoT and Spitzer  
and a $\sim$5-yr RV monitoring with HARPS, we updated the CoRoT-9b physical parameters,
$\mp=0.84\pm0.05~\Mjup$, $\rp=1.066^{+0.075}_{-0.063}~\Rjup$, and 
$\rho_{\rm p}=0.86^{+0.18}_{-0.16}~\rm g\;cm^{-3}$,  
which agree well with the literature values (D10). 
With the new RV data, we found a higher significance 
for the non-zero eccentricity $e=0.133^{+0.042}_{-0.037}$
of CoRoT-9b and no evidence for additional companions. 
The TTVs do not deviate significantly 
from a linear ephemeris either, thus supporting the absence of a strong perturber,
even though only four transit epochs could be determined.
Inner transiting (coplanar) planets with $\rp \gtrsim 1.3$, 2.0, 2.3, and 2.6~$\Rearth$ 
and orbital periods $P=1$, 10, 25, and 50~d, respectively, 
can be excluded as well using CoRoT data.

We investigated different scenarios to reproduce the orbit of CoRoT-9b, assuming a lack of additional companions. 
We argue that in situ formation, secular interactions with an outer perturber, 
and single planet migration are all unlikely to reproduce the CoRoT-9 system.  
Instead we believe that planet-planet scattering is the most likely origin story for the system.

Using simulations of planet-planet scattering we showed that the eccentricity of CoRoT-9b  is likely a residual scar 
from an instability that ejected a planet of roughly $50~\Mearth$.  
Although we only performed simulations with a single additional planet, 
we expect that the same process could have occurred with more planets. 
The ejection of multiple planets with a total of $\sim 50~\Mearth$ could induce a comparable eccentricity in CoRoT-9b.

Our simulations also serve to motivate future observations of the Rossiter-McLaughlin effect of CoRoT-9b, 
although challenging given the long transit duration of 8.3~hr \citep[but see e.g.][for the case of HD\,80606\,b]{2010A&A...516A..95H}. 
While planet-planet scattering of a $\sim50~\Mearth$ planet can excite the eccentricity of CoRoT-9b to its current value, there is little associated inclination excitation.  The orbital plane of CoRoT-9b should not have changed since before the instability.  We thus predict a spin-orbit alignment for the system.  Yet perhaps the most interesting outcome would be a measured misalignment, which would be strong indirect evidence that the orbital plane of CoRoT-9b was already misaligned before the instability, perhaps indicating that the protoplanetary disk of the star was tilted at a young age.

\bibliographystyle{aa} 
\bibliography{Bonomoetal_CoRoT-9} 

\begin{acknowledgements}
This work is based in part on observations made with the \emph{Spitzer} Space Telescope, which is operated by the Jet Propulsion Laboratory, California Institute of Technology under a contract with NASA.
A.S.B. acknowledges funding from the European Union Seventh Framework programme (FP7/2007-2013) 
under grant agreement No. 313014 (ETAEARTH). 
G.H. and A.L. acknowledge the support by the CNES and the French Agence Nationale de la Recherche (ANR), 
under programme ANR-12-BS05-0012 ``Exo-Atmos''.
S.N.R. and A.I. thank the Agence Nationale pour la Recherche for support via grant ANR-13-BS05-0003- 01 (project MOJO). A.I. also thanks partial financial support from FAPESP (Proc. 16/12686-2 and 16/19556-7).
R.A. acknowledges the Spanish Ministry of Economy and Competitiveness (MINECO) for the financial support under the Ram\'on y Cajal programme 
RYC-2010-06519, and under the programme RETOS ESP2014-57495-C2-1-R and ESP2016-80435-C2-2-R. 
H.D. acknowledges support by grant ESP2015-65712-C5-4-R of the Spanish Secretary of State for R\&D\&i (MINECO).
\end{acknowledgements}

\end{document}